\documentclass[secnumarabic, prb]{revtex4}
\usepackage{bm}
\usepackage{amsmath2000}
\usepackage{hyperref}

\newcommand{\be}{\begin{equation}}
\newcommand{\ee}{\end{equation}}
\newcommand{\lb}{\label}

\begin{document}

\title{Gravitation and electromagnetism}
\author{Valery P. Dmitriyev}
\affiliation{Lomonosov University\\
P.O.Box 160, Moscow 117574, Russia} \email{dmitr@cc.nifhi.ac.ru}
\date{23 July 2002}

\begin{abstract}
Maxwell's equations comprise both electromagnetic and
gravitational fields. The transverse part of the vector potential
belongs to magnetism, the longitudinal one is concerned with
gravitation. The Coulomb gauge indicates that longitudinal
components of the fields propagate instantaneously. The
delta-function singularity of the field of the divergence of the
vector potential, referred to as the dilatation center, represents
an elementary agent of gravitation. Viewing a particle as a source
or a scattering center of the point dilatation, the Newton's
gravitation law can be reproduced.

\end{abstract}

\maketitle
\section{Maxwell's equations in the Kelvin-Helmholtz
representation}

      The general form of Maxwell's equations is given by
\be \frac{1}{\,c}\,\frac{\partial {\bf A}}{\partial t} + {\bf E} +
\nabla \varphi  = 0 \, ,\tag{1.1}\lb{1.1}\ee \be \frac{\partial
{\bf E}}{\partial t} - c\,\nabla  \times \nabla \times {\bf A} +
4\pi {\bf j} = 0 \, ,\tag{1.2}\lb{1.2}\ee \be \nabla\!  \cdot {\bf
E} = 4\pi \rho \, .\tag{1.3}\lb{1.3}\ee The Helmholtz theorem: a
vanishing at infinity vector field ${\bf u}$ can be expanded into
the sum of solenoidal ${\bf u}_{\rm r}$ and potential ${\bf
u}_{\rm g}$ components. We have for the electric field: \be {\bf
E} = {\bf E}_{\rm r}  + {\bf E}_{\rm g} \, ,\tag{1.4}\lb{1.4}\ee
where \be \nabla  \cdot {\bf E}_{\rm r} = 0 \,
,\tag{1.5}\lb{1.5}\ee \be \nabla\! \times {\bf E}_{\rm g} = 0 \,
.\tag{1.6}\lb{1.6}\ee The respective expansion for the vector
potential can be written as \be {\bf A} = {\bf A}_{\rm r} +
\frac{c}{\,\,c_{\rm g}}\,\, {\bf A}_{\rm g} \,
,\tag{1.7}\lb{1.7}\ee where \be \nabla \! \cdot {\bf A}_{\rm r}  =
0 \, ,\tag{1.8}\lb{1.8}\ee \be \nabla \times {\bf A}_{\rm g}  = 0
\, ,\tag{1.9}\lb{1.9}\ee and $c_{\rm g}$ is a constant. Substitute
(1.4) and (1.7) into (1.1): \be \frac{1}{c}\,\frac{\partial {\bf
A}_{\rm r}}{\partial t} + {\bf E}_{\rm r} + \frac{1}{\,\,c_{\rm
g}}\frac{\partial {\bf A}_{\rm g}}{\partial t} + {\bf E}_{\rm g} +
\nabla \varphi  = 0 \, .\tag{1.10}\lb{1.10}\ee Taking the curl of
(1.10), we get through (1.6) and (1.9) \be \nabla \times \left(
{\frac{1}{\,c}\,\frac{\partial {\bf A}_{\rm r}}{\partial t} + {\bf
E}_{\rm r}} \right) = 0 \, .\tag{1.11}\lb{1.11}\ee On the other
hand, by (1.5) and (\ref{1.8}) we have \be \nabla  \cdot \left(
{\frac{1}{\,c}\,\frac{\partial {\bf A}_{\rm r}}{\partial t} + {\bf
E}_{\rm r}} \right) = 0 \, .\tag{1.12}\lb{1.12}\ee If the
divergence and curl of a field equal to zero, then the very field
is vanishing. Hence (\ref{1.11}) and (1.12) imply that \be
\frac{1}{\,c}\,\frac{\partial {\bf A}_{\rm r}}{\partial t} + {\bf
E}_{\rm r} = 0 \, .\tag{1.13}\lb{1.13}\ee Subtracting (1.13) from
(\ref{1.10}) we get also \be \frac{1}{\,\,c_{\rm
g}}\,\frac{\partial {\bf A}_{\rm g}}{\partial t} + {\bf E}_{\rm g}
+ \nabla \varphi = 0 \, .\tag{1.14}\lb{1.14}\ee Similarly,
expanding as well the density of the current \be {\bf j} = {\bf
j}_{\rm r} + {\bf j}_{\rm g} \, ,\tag{1.15}\lb{1.15}\ee \be
\nabla\! \cdot {\bf j}_{\rm r} = 0 \, ,\tag{1.16}\lb{1.16}\ee \be
\nabla \times {\bf j}_{\rm g}  = 0 \, ,\tag{1.17}\lb{1.17}\ee
(\ref{1.2}) can be broken up in two equations \be \frac{\partial
{\bf E}_{\rm r}}{\partial t}  - c\,\nabla  \times \nabla  \times
{\bf A}_{\rm r}  + 4\pi {\bf j}_{\rm r}  = 0 \,
,\tag{1.18}\lb{1.18}\ee \be \frac{\partial {\bf E}_{\rm
g}}{\partial t}  + 4\pi {\bf j}_{\rm g}  = 0 \,
.\tag{1.19}\lb{1.19}\ee Through (1.4) and (1.5) equation
(\ref{1.3}) will be \be \nabla \cdot {\bf E}_{\rm g}  = 4\pi \rho
\, .\tag{1.20}\lb{1.20}\ee

\section{Wave equations}

Let us derive from (1.13), (1.14), (1.18), (1.19) and (1.20) the
wave equations for the solenoidal (transverse) and potential
(longitudinal) components of the fields. In what follows we will
use the general vector relation \be \nabla \left( {\nabla  \cdot
{\bf u}} \right) = \nabla ^2 {\bf u} + \nabla  \times \nabla
\times {\bf u} \, .\tag{2.1}\lb{2.1}\ee The wave equation for
${\bf A}_{\rm r}$ can be found thus. Differentiate (1.13) with
respect to time: \be \frac{1}{\,c}\,\frac{\partial ^2\!{\bf
A}_{\rm r}}{\partial t^2}  + \frac{\partial {\bf E}_{\rm
r}}{\partial t} = 0 \, .\tag{2.2}\lb{2.2}\ee Substitute (1.18)
into (2.2). With the account of (2.1) we get \be \frac{\partial
^2\!{\bf A}_{\rm r}}{\partial t^2}  -\, c^2 \nabla ^2 {\bf A}_{\rm
r} = 4\pi c{\bf j}_{\rm r} \, .\tag{2.3}\lb{2.3}\ee The wave
equation for ${\bf E}_{\rm r}$   can be found as follows.
Differentiate (1.18) with respect to time \be \frac{\partial
^2\!{\bf E}_{\rm r}}{\partial t^2}   - c\,\nabla \times \nabla
\times \frac{\partial {\bf A}_{\rm r}}{\partial t} + 4\pi
\frac{\partial {\bf j}_{\rm r}}{\partial t} = 0 \,
.\tag{2.4}\lb{2.4}\ee Substitute (\ref{1.13}) into (\ref{2.4}).
With the account of (2.1) we get \be \frac{\partial ^2{\bf E}_{\rm
r}}{\partial t^2}
 - c^2 \nabla ^2 {\bf E}_{\rm r} =  -\, 4\pi
c\partial _t {\bf j}_{\rm r} \, .\tag{2.5}\lb{2.5}\ee In order to
find the wave equations for the potential fields we need a gauge
relation. Let us postulate for the potential part of the vector
potential the specific Lorentz gauge \be \nabla\!  \cdot {\bf
A}_{\rm g}  + \frac{1}{\,\,{c}_{\rm g}}\, \frac{\partial
\varphi}{\partial t} = 0 \, ,\tag{2.6}\lb{2.6}\ee where in general
$c_{\rm g} \ne c$. The solenoidal part of the vector potential
meets automatically the Coulomb gauge (\ref{1.8}). The wave
equation for ${\bf A}_{\rm g}$ can be found as follows.
Differentiate (\ref{1.14}) with respect to time: \be
\frac{1}{\,\,c_{\rm g}}\,\frac{\partial ^2{\bf A}_{\rm
g}}{\partial t^2} + \frac{\partial {\bf E}_{\rm g}}{\partial t} +
\frac{\partial \nabla \varphi}{\partial t} = 0 \,
.\tag{2.7}\lb{2.7}\ee Take the gradient of (2.6): \be \nabla
\left( {\nabla\!  \cdot {\bf A}_{\rm g} } \right) +
\frac{1}{\,\,c_{\rm g}}\, \nabla \frac{\partial \varphi}{\partial
t}  = 0 \, .\tag{2.8}\lb{2.8}\ee Combine (2.7), (2.8) and
(\ref{1.19}). With the account of (\ref{2.1}) we get \be
\frac{\partial ^2\!{\bf A}_{\rm g}}{\partial t^2}  - c_g^2 \nabla
^2 {\bf A}_{\rm g} = 4\pi c_{\rm g}{\bf j}_{\rm g} \,
.\tag{2.9}\lb{2.9}\ee Next, we will find the wave equation for
$\varphi$. Take the divergence of (\ref{1.14}): \be
\frac{1}{\,\,c_{\rm g}}\,\frac{\partial \nabla\! \cdot\! {\bf
A}_{\rm g}}{\partial t} + \nabla\! \cdot {\bf E}_{\rm g} + \nabla
^2 \varphi = 0 \, .\tag{2.10}\lb{2.10}\ee Combine (2.10), (2.6)
and (\ref{1.20}): \be \frac{\partial ^2\varphi}{\partial t^2}
  - c_g^2 \nabla ^2 \varphi  = 4\pi c_{\rm g}^{\rm 2} \rho
\, .\tag{2.11}\lb{2.11}\ee The wave equation for ${\bf E}_{\rm g}$
we will find from the wave equations of ${\bf A}_{\rm g}$  and
$\varphi$, using (\ref{1.14}). Differentiate  (2.9) with respect
to time \be \frac{\partial ^2}{\partial t^2} \frac{\partial{\bf
A}_{\rm g}}{\partial t}  - c_g^2 \nabla ^2 \frac{\partial{\bf
A}_{\rm g}}{\partial t} = 4\pi c_{\rm g}\frac{\partial{\bf j}_{\rm
g}}{\partial t} \, .\tag{2.12}\lb{2.12}\ee Take the gradient of
(2.11) \be \frac{\partial ^2\nabla \varphi}{\partial t^2}   -
c_g^2 \nabla ^2 \nabla \varphi  = 4\pi c_{\rm g}^2\nabla \rho \,
.\tag{2.13}\lb{2.13}\ee Summing (2.12) and (2.13), we get with the
account of (\ref{1.14}) \be \frac{\partial ^2{\bf E}_{\rm
g}}{\partial t^2}  -\, c_g^2 \nabla ^2 {\bf E}_{\rm g} =  -\, 4\pi
\left( c_g^2 \nabla \rho  + \frac{\partial{\bf j}_{\rm
g}}{\partial t} \right) \, .\tag{2.14}\lb{2.14}\ee Thus, Maxwell's
equations (\ref{1.1})-(1.3) with the specific Lorentz gauge
(\ref{2.6}) imply that the solenoidal and potential components of
the fields propagate with different velocities. Solenoidal
components propagate with the speed $c$ of light. Their wave
equations are (\ref{2.3}) and (\ref{2.5}). Potential components
and the electrostatic potential propagate with a speed $c_{\rm
g}$. Their wave equations are (2.9), (2.11) and (2.14).

\section{Quasielasticity}

Equations (\ref{2.3}) and (\ref{2.9}) have the character of the
elastic equations. In this connection, the vector potential ${\bf
A}$ can be correspondent\cite{1} with a certain displacement field
${\bf s}$, and the density ${\bf j}$ of the current -- with the
density ${\bf f}$ of an external force . The gauge relation
(\ref{2.6}) is interpreted as a linearized continuity equation, in
which the constant $c_{\rm g}$  has directly the meaning of the
speed of an expansion-contraction wave \cite{2}. We are interested
in the interaction of two external forces ${\bf f}_1$ and  ${\bf
f}_2$, which produce elastic fields ${\bf s}_1$  and  ${\bf s}_2$,
respectively. The energy of the elastic interaction is given by
the general relation \be U_{12}  =  - \,\varsigma \int {{\bf f}_1
\cdot {\bf s}} _2d^3 x = - \,\varsigma \int {{\bf f}_2  \cdot {\bf
s}} _1d^3 x \, ,\tag{3.1}\lb{3.1}\ee where the sign minus in (3.1)
corresponds to conditions of the Clapeyron theorem \cite{3},
$\varsigma$  is a constant.

The energy of the static interaction can be found substituting
into (3.1) \be {\bf s} \sim \frac{1}{\,c}\,{\bf A} \,
,\tag{3.2}\lb{3.2}\ee \be {\bf f} \sim 4\pi c{\bf j} \,
,\tag{3.3}\lb{3.3}\ee \be \varsigma \sim \frac{1}{\,4\pi c} \,
.\tag{3.4}\lb{3.4}\ee

We have for the transverse interaction \be U_{\rm r}  =  -
\frac{1}{c}\int {{\bf j}_{\rm r}  \cdot {\bf A}} _{\rm r}d^3 x \,
.\tag{3.5}\lb{3.5}\ee Suppose that \be c_{\rm g}  >  > c \,
.\tag{3.6}\lb{3.6}\ee Then, through (\ref{1.7}) relations
(\ref{2.6}) and (\ref{1.8}) turn to the Coulomb gauge \be \nabla\!
\cdot {\bf A} = 0 \, .\tag{3.7}\lb{3.7}\ee We have according to
(\ref{1.16}) and (1.17) \be {\bf j}_{\rm r}  = \nabla \times {\bf
R},\quad {\bf j}_{\rm g}  = \nabla G \, ,\tag{3.8}\lb{3.8}\ee
where ${\bf R}$ and $G$   are vector and scalar fields. Using
(\ref{1.15}), (3.8), (3.7), (1.7) and (\ref{1.9}) take the
following integral by parts: \be
\begin{array}{l}
 \int {{\bf j} \cdot {\bf A}} d^3 x = \int {\left( {{\bf j}_{\rm r}  + \nabla G} \right) \cdot {\bf A}} d^3 x = \int {{\bf j}_{\rm r}  \cdot {\bf A}} d^3 x \\
\\
  = \int {\left( {\nabla\!  \times {\bf R}} \right) \cdot ( {{\bf A}_{\rm r} + c\, {\bf A}_{\rm g}}/c_{\rm g} )} d^3 x = \int {{\bf j}_{\rm r}  \cdot {\bf A}} _{\rm r}d^3 x \, . \\
 \end{array}
\tag{3.9}\lb{3.9}\ee From (3.9) and (3.5) we get the regular
expression for the energy of magnetostatic interaction \be U_{\rm
r}  =  - \frac{1}{\,c}\int {{\bf j} \cdot {\bf A}} d^3 x \,
.\tag{3.10}\lb{3.10}\ee Elementary sources of the magnetic field
correspond to the two forms of the external force density ${\bf
f}$ (3.3). The point force at ${\bf x'}$: \be {\bf f} = 4\pi
cq{\bf v}\delta \left( {{\bf x} - {\bf x'}} \right) \,
,\tag{3.11}\lb{3.11}\ee and the torsion center at ${\bf x'}$: \be
{\bf f}_{\rm r}  = 4\pi ca\nabla  \times \left[ {{\bm{\mu} }\delta
\left( {{\bf x} - {\bf x'}} \right)}
\right]\,,\tag{3.12}\lb{3.12}\ee where $q{\bf v}$  and $a{\bm
{\mu} }$ are constant vectors, $\left| {\bm {\mu} } \right| = 1$.
They describe a moving electric charge and a point magnetic
dipole, respectively \cite{1}. Substituting (3.11) and
(\ref{3.12}) into the right-hand part of the equation (\ref{2.3})
we can find the fields $ {\bf A}$ produced by these forces. Then,
substituting these fields into (\ref{3.10}) and (\ref{3.5}), we
arrive at the well-known expressions for the interaction energies
of electric currents and point magnetic dipoles.

The elementary source of the longitudinal part ${\bf A}_{\rm g}$
  of the vector potential is given by the density of the external force of the
form \be {\bf f}_{\rm g}  =  -\, 4\pi c_{\rm g} b\nabla \delta
\left( {{\bf x} - {\bf x'}} \right) \, ,\tag{3.13}\lb{3.13}\ee
where $b$ is the strength of the dilatation center (3.13)
\cite{4}. Substitute (3.13) into the right-hand part of the static
variant of the equation (\ref{2.9}): \be c_{\rm g} \nabla ^2 {\bf
A}_{\rm g} = 4\pi b\nabla \delta \left( {{\bf x} - {\bf x'}}
\right) \, .\tag{3.14}\lb{3.14}\ee With the account of (\ref{2.1})
and (\ref{1.9}) we get from (3.14) \be c_{\rm g} \nabla \! \cdot
{\bf A}_{\rm g} = 4\pi b\delta \left( {{\bf x} - {\bf x'}} \right)
\, .\tag{3.15}\lb{3.15}\ee Following (\ref{3.1})-(3.4) we have for
the energy of longitudinal interaction: \be U_{\rm g}  =  -
\frac{1}{\,\,{c}_{\rm g}} \int {{\bf j}_{\rm g} \cdot {\bf A}}
_{\rm g}d^3 x \, .\tag{3.16}\lb{3.16}\ee Substitute (3.13) with
the account of (\ref{3.3}) into (3.16): \be U_{{\rm 12}}  =
\frac{b_1}{c_{\rm g}} \int {\nabla \delta \left( {{\bf x} - {\bf
x}_{\bf 1} } \right) \cdot {\bf A}} _{\bf 2}d^3 x =  -\,
\frac{b_1}{c_{\rm g}} \int {\delta \left( {{\bf x} - {\bf x}_{\bf
1} } \right)\nabla\! \cdot {\bf A}} _{\bf 2}d^3 x \,
.\tag{3.17}\lb{3.17}\ee Substituting (3.15) into (3.17), we get
\be U_{{\rm 12}}  =  -\, \frac{4\pi b_1 b_2}{c_{\rm g}^{\rm 2}}
\int\! {\delta \left( {{\bf x} - {\bf x}_{\bf 1} } \right)\delta
\left( {{\bf x} - {\bf x}_{\bf 2} } \right)} d^3 x \,
\tag{3.18}\lb{3.18}\ee \be
 =  -\, \frac{4\pi b_1 b_2}{c_{\rm g}^{\rm 2}}\,\, \delta \left( {{\bf x}_{\bf 1}  - {\bf x}_{\bf 2} } \right)
\, .\tag{3.19}\lb{3.19}\ee Expression (3.19) implies, that two
dilatation centers (3.15) interact with each other only if they
are in a direct contact. The sign of (3.18), or (3.19), indicates
that this is the attraction.

Take notice that solenoidal and potential fields are orthogonal to
each other in the sense of  (\ref{3.1}). Indeed, using
(\ref{3.8}), (\ref{1.8}) and (1.9), we find that \be \int {{\bf
j}_{\rm g} \cdot {\bf A}} _{\rm r}d^3 x = \int {\nabla G \cdot
{\bf A}_{\rm r}} d^3 x = - \int {G\nabla\!  \cdot {\bf A}_{\rm r}
} d^3 x = 0 \, ,\tag{3.20}\lb{3.20}\ee \be \int {{\bf j}_{\rm r}
\cdot {\bf A}} _{\rm g}d^3 x = \int {\nabla\!  \times {\bf R}
\cdot {\bf A}_{\rm g}} d^3 x = \int {{\bf R} \cdot \nabla \times
{\bf A}_{\rm g} } d^3 x = 0 \, .\tag{3.21}\lb{3.21}\ee

\section{Gravitation}

We consider dilatation centers distributed with the volume density
$bp\left( {\bf x} \right)$. Then equation (\ref{3.15}) becomes \be
c_{\rm g} \nabla \cdot {\bf A}_{\rm g} = 4\pi bp\left( {\bf x}
\right) \, .\tag{4.1}\lb{4.1}\ee The interaction energy of the two
clusters, or clouds, of dilatation centers can be found
substituting delta-functions in (3.18) by the reduced densities
$p\left( {\bf x} \right)$   of the distributions. This gives \be
U_{{\rm 12}} =  -\, \frac{4\pi b_1 b_2}{c_{\rm g}^{\rm 2}} \int\!
{p_1 \left( {\bf x} \right)p_2 \left( {\bf x} \right)} d^3 x \,
.\tag{4.2}\lb{4.2}\ee Consider a weak source at ${\bf x}^*$, which
emits dilatation centers with a sufficiently high linear velocity
$\upsilon _{\rm g}$. Such a source will create a quick-formed
stationary distribution of the point dilatation with the reduced
density \be p\left( {\bf x} \right) = \frac{g}{{4\pi \upsilon
_{\rm g} \left( {{\bf x} - {\bf x^*}} \right){}^2}} \,
,\tag{4.3}\lb{4.3}\ee where  $g$ is a universal constant.
Substituting (4.3) into (\ref{4.2}), we find the interaction
energy for two sources of the point dilatation \be U_{12}  =  -\,
\frac{{g^2 b_1 b_2 }}{{4\pi c_{\rm g}^{\rm 2} \upsilon _{\rm
g}^{\rm 2} }}\int\! {\frac{{d^3 x}}{{\left( {{\bf x} - {\bf
x}_{\bf 1} } \right)^2 \left( {{\bf x} - {\bf x}_{\bf 2} }
\right)^2 }}} = -\, \frac{{\pi ^2 g^2 }}{{4c_{\rm g}^{\rm 2}
\upsilon _{\rm g}^{\rm 2} }}\,\,\frac{{b_1 b_2 }}{{\left| {{\bf
x}_{\bf 1}  - {\bf x}_{\bf 2} } \right|}} \, .\tag{4.4}\lb{4.4}\ee
We will assume that each particle is a weak source of the point
dilatation (\ref{3.15}) or a scattering center in a dynamic sea of
the point dilatation, the strength $b$ of the source   being
proportional to the particle's mass.  Then relation  (4.4) will be
a model of the Newton's law of gravitation.

Notice that in the model thus constructed we must distinguish the
speed  $\upsilon _{\rm g}$, with which the gravitational
interaction is transmitted, and the speed $c_{\rm g}$  of the
longitudinal wave.  The latter can be interpreted as the
gravitational wave.

Thus, gravitation enters into the general structure of Maxwell's
equations.  A gravitating center is formally modeled by a
potential component of the current having the form \be {\bf
j}_{\rm g}  =  - \frac{{gb}}{{4\pi \upsilon _{\rm g} }}\nabla
\frac{1}{{\left( {{\bf x} - {\bf x}^ *  } \right)^2
}}\,.\tag{4.5}\lb{4.5}\ee And the gravitational interaction is
calculated by means of the general relation (\ref{3.16}), where
the longitudinal component ${\bf A}_{\rm g}$ of the vector
potential is found substituting (4.5) into the longitudinal part
(\ref{2.9}) of Maxwell's equations.

\section{Conclusion}

Maxwell's equations (\ref{1.1})-(1.3) describe both
electromagnetic and gravitational fields. The transverse part of
the vector potential belongs to magnetism, and longitudinal one is
concerned with gravitation. Transverse fields propagate with the
speed of light. The Coulomb gauge (\ref{3.7}) indicates that
longitudinal waves propagate in effect instantaneously, comparing
with transverse waves. Choosing properly expressions for the
current density, magnetic and gravitational interactions can be
modeled. An elementary agent of the gravitational interaction
corresponds to the dilaton, which is a delta-function singularity
(\ref{3.15}) of the field of the divergence of the vector
potential . The sources of longitudinal and transverse fields do
not interact with each other. This signifies that gravitation can
not be detected with the aid of light.

In the end it should be noted that some of the questions
considered here and in \cite{1} were recently approached in
\cite{5}.

\end{document}